\documentclass{article}

\input{tcilatex}
\begin{document}

\author{Nick Laskin}
\title{\textbf{New Polynomials and Numbers Associated with Fractional
Poisson Probability Distribution}}
\date{\textit{TopQuark Inc., Toronto, ON, M6P 2P2, Canada}}
\maketitle

\begin{abstract}
Generalizations of Bell polynomials, Bell numbers, and Stirling numbers of
the second kind have been introduced and their generating functions were
evaluated.


\end{abstract}

\section{Fractional Poisson Probability Distribution}

Bell polynomials, Bell numbers \cite{Bell} and Stirling numbers of the
second kind \cite{Stirling}-\cite{Charalambides} are related to the well
known Poisson probability distribution. Recently, the fractional Poisson
probability distribution has been developed \cite{Laskin1}, and some of its
applications have been implemented in \cite{Laskin2}. The fractional Poisson
probability distribution is a natural model which captures long memory
impact on the counting process.

Here we present and explore generating functions for new polynomials and
numbers related to the fractional Poisson probability distribution. Those
new polynomials and numbers are: fractional Bell polynomials, fractional
Bell numbers and fractional Stirling numbers of the second kind (see, \cite%
{Laskin2}).

The fractional Poisson probability distribution $P_{\mu }(n,t)$ of arriving $%
n$ items ($n=0,1,2,...$) by time $t$ is given by \cite{Laskin1}

\begin{equation}
P_{\mu }(n,t)=\frac{(\nu t^{\mu })^{n}}{n!}\sum\limits_{k=0}^{\infty }\frac{%
(k+n)!}{k!}\frac{(-\nu t^{\mu })^{k}}{\Gamma (\mu (k+n)+1))},\qquad 0<\mu
\leq 1,  \label{eq1}
\end{equation}

where the parameter $\nu $ has physical dimension $[\nu ]=\sec ^{-\mu }$ and
the gamma function $\Gamma (\mu )$ has the familiar representation $\Gamma
(\mu )=\int\limits_{0}^{\infty }dte^{-t}t^{\mu -1}$, $\mathrm{Re}\mu >0$. At 
$\mu =1$ Eq.(\ref{eq1}) is transformed into the well known equation for the
standard Poisson probability distribution with substitution $\nu \rightarrow 
\overline{\nu }$, where $\overline{\nu }$ is the rate of arrivals of the
standard Poisson process with physical dimension $\overline{\nu }=\sec ^{-1}$%
.

\section{Fractional Bell polynomials and Bell numbers}

Based on the fractional Poisson probability distribution Eq.(\ref{eq1}), we
introduce a new generalization of Bell polynomials

\begin{equation}
B_{\mu }(x,m)=\sum\limits_{n=0}^{\infty }n^{m}\frac{x^{n}}{n!}%
\sum\limits_{k=0}^{\infty }\frac{(k+n)!}{k!}\frac{(-x)^{k}}{\Gamma (\mu
(k+n)+1))},\qquad B_{\mu }(x,0)=1,  \label{eq2}
\end{equation}

where the parameter $\mu $ is $0<\mu \leq 1.$ We will call $B_{\mu }(x,m)$
as fractional Bell polynomials. Polynomials $B_{\mu }(x,m)$ are related to
the well-known Bell polynomials \cite{Bell} $B(x,m)$ by $B_{\mu }(x,m)|_{\mu
=1}=B(x,m)=e^{-x}\sum\limits_{n=0}^{\infty }n^{m}\frac{x^{n}}{n!}.$ From Eq.(%
\ref{eq2}) we come to the formula for numbers $B_{\mu }(m)$, which we call
the fractional Bell numbers

\begin{equation}
B_{\mu }(m)=B_{\mu }(x,m)|_{x=1}=\sum\limits_{n=0}^{\infty }\frac{n^{m}}{n!}%
\sum\limits_{k=0}^{\infty }\frac{(k+n)!}{k!}\frac{(-1)^{k}}{\Gamma (\mu
(k+n)+1))}.  \label{eq39}
\end{equation}

Now we focus on the general definitions given by Eqs.(\ref{eq2}) and (\ref%
{eq39}) to find the generating functions of polynomials $B_{\mu }(x,m)$ and
numbers $B_{\mu }(m).$ Let us introduce the generating function $F_{\mu
}(s,x)$ of polynomials $B_{\mu }(x,m)$ as 
\begin{equation}
F_{\mu }(s,x)=\sum\limits_{m=0}^{\infty }\frac{s^{m}}{m!}B_{\mu }(x,m).
\label{eq42}
\end{equation}

To find an explicit equation for $F_{\mu }(s,x)$, we substitute Eq.(\ref{eq2}%
) into Eq.(\ref{eq42}) and evaluate the sum over $m$. As a result we have

\begin{equation}
F_{\mu }(s,x)=E_{\mu }(x(e^{s}-1)),  \label{eq44}
\end{equation}

where $E_{\mu }(z)$ is the Mittag-Leffler function given by its power series 
\cite{Erdelyi}

\begin{equation}
E_{\mu }(z)=\sum\limits_{m=0}^{\infty }\frac{z^{m}}{\Gamma (\mu m+1)}.
\label{eq6}
\end{equation}

If we put $x=1$ in Eq.(\ref{eq44}), then we immediately come to the
generating function $\mathcal{B}_{\mu }(s)$ of the fractional Bell numbers $%
B_{\mu }(m)$

\begin{equation}
\mathcal{B}_{\mu }(s)=\sum\limits_{m=0}^{\infty }\frac{s^{m}}{m!}B_{\mu
}(m)=E_{\mu }(e^{s}-1).  \label{eq47}
\end{equation}

In the case of $\mu =1$, Eq.(\ref{eq44}) turns into the equation for the
generating function of the Bell polynomials $F_{1}(s,x)=\exp \{x(e^{s}-1)\},$
while Eq.(\ref{eq47}) reads $\mathcal{B}_{1}(s)=\sum\limits_{m=0}^{\infty }%
\frac{s^{m}}{m!}B_{1}(m)=\exp (e^{s}-1),$ and we come to the equation for
the Bell numbers generating function.

\section{Fractional Stirling numbers of the second kind}

Now we introduce the fractional generalization of the Stirling numbers of
the second kind $S_{\mu }(m,l)$ by means of equation

\begin{equation}
B_{\mu }(x,m)=\sum\limits_{l=0}^{m}S_{\mu }(m,l)x^{l},  \label{eq55}
\end{equation}

where $B_{\mu }(x,m)$ is a fractional generalization of Bell polynomials
given by Eq.(\ref{eq2}) and the parameter $\mu $ is $0<\mu \leq 1$. At $\mu
=1$, Eq.(\ref{eq55}) defines integers $S(m,l)=S_{\mu }(m,l)|_{\mu =1}$,
which are called Stirling numbers of the second kind. At $x=1$, when
fractional Bell polynomials $B_{\mu }(x,m)$ become fractional Bell numbers, $%
B_{\mu }(m)=B_{\mu }(x,m)|_{x=1}$, Eq.(\ref{eq55}) gives us the equation to
express fractional Bell numbers in terms of fractional Stirling numbers of
the second kind $B_{\mu }(m)=\sum\limits_{l=0}^{m}S_{\mu }(m,l).$

To find $S_{\mu }(m,l)$ we transform the right-hand side of Eq.(\ref{eq2})
as follows

\begin{equation}
B_\mu (x,m)=\sum\limits_{l=0}^\infty \frac{x^l}{\Gamma (\mu l+1)}%
\sum\limits_{n=0}^l(-1)^{l-n}\binom lnn^m,  \label{eq59}
\end{equation}

where the notation$\binom{l}{n}=\frac{l!}{n!(l-n)!}$ has been introduced. By
comparing Eq.(\ref{eq55}) and Eq.(\ref{eq59}) we conclude that the
fractional Stirling numbers $S_{\mu }(m,l)$ are given by 
\begin{equation}
S_{\mu }(m,l)=\frac{1}{\Gamma (\mu l+1)}\sum\limits_{n=0}^{l}(-1)^{l-n}%
\binom{l}{n}n^{m},  \label{eq60}
\end{equation}

\[
S_{\mu }(m,0)=\delta _{m,0},\qquad S_{\mu }(m,l)=0,\quad l=m+1,\quad
m+2,.... 
\]

As an example, Table 1 presents a few fractional Stirling numbers of the
second kind.

\begin{tabular}{|l|l|l|l|l|l|l|}
\hline
m\textit{\TEXTsymbol{\backslash}l} & $1$ & $2$ & $3$ & $4$ & $5$ & $6$ \\ 
\hline
1 & $\frac{1}{\Gamma (\mu +1)}$ &  &  &  &  &  \\ \hline
2 & $\frac{1}{\Gamma (\mu +1)}$ & $\frac{2}{\Gamma (2\mu +1)}$ &  &  &  & 
\\ \hline
3 & $\frac{1}{\Gamma (\mu +1)}$ & $\frac{6}{\Gamma (2\mu +1)}$ & $\frac{6}{%
\Gamma (3\mu +1)}$ &  &  &  \\ \hline
4 & $\frac{1}{\Gamma (\mu +1)}$ & $\frac{14}{\Gamma (2\mu +1)}$ & $\frac{36}{%
\Gamma (3\mu +1)}$ & $\frac{24}{\Gamma (4\mu +1)}$ &  &  \\ \hline
5 & $\frac{1}{\Gamma (\mu +1)}$ & $\frac{30}{\Gamma (2\mu +1)}$ & $\frac{150%
}{\Gamma (3\mu +1)}$ & $\frac{240}{\Gamma (4\mu +1)}$ & $\frac{120}{\Gamma
(5\mu +1)}$ &  \\ \hline
6 & $\frac{1}{\Gamma (\mu +1)}$ & $\frac{62}{\Gamma (2\mu +1)}$ & $\frac{540%
}{\Gamma (3\mu +1)}$ & $\frac{1560}{\Gamma (4\mu +1)}$ & $\frac{1800}{\Gamma
(5\mu +1)}$ & $\frac{720}{\Gamma (6\mu +1)}$ \\ \hline
\end{tabular}

Table 1. \textit{Fractional Stirling numbers of the second kind }$S_{\mu
}(m,l)$\textit{\ }($0<\mu \leq 1$)

To find a generating function of the fractional Stirling numbers $S_{\mu
}(m,l)$, let's substitute $B_{\mu }(x,m)$ from Eq.(\ref{eq55}) into the
definition given by Eq.(\ref{eq42}). Hence, we have

\begin{equation}
F_\mu (s,x)=\sum\limits_{m=0}^\infty \frac{s^m}{m!}\left(
\sum\limits_{l=0}^mS_\mu (m,l)x^l\right) =\sum\limits_{l=0}^\infty \left(
\sum\limits_{m=l}^\infty S_\mu (m,l)\frac{s^m}{m!}\right) x^l.  \label{eq67}
\end{equation}

On the other hand, from Eq.(\ref{eq44}), we have $F_\mu
(s,x)=\sum\limits_{l=0}^\infty \frac{(e^s-1)^l}{\Gamma (\mu l+1)}x^l$. Upon
comparing this equation and Eq.(\ref{eq67}) we introduce the generating
function $\mathcal{G}_\mu (s,l)$

\begin{equation}
\mathcal{G}_\mu (s,l)=\sum\limits_{m=l}^\infty S_\mu (m,l)\frac{s^m}{m!}=%
\frac{(e^s-1)^l}{\Gamma (\mu l+1)},\qquad l=0,1,2,....  \label{eq69}
\end{equation}

In addition, we introduce the generating function $\mathcal{F}_\mu (s,t)$
defined by

\begin{equation}
\mathcal{F}_\mu (s,t)=\sum\limits_{m=0}^\infty \sum\limits_{l=0}^mS_\mu (m,l)%
\frac{s^mt^l}{m!}=\sum\limits_{l=0}^\infty \frac{t^l(e^s-1)^l}{\Gamma (\mu
l+1)}=E_\mu (t(e^s-1)).  \label{eq71}
\end{equation}

As a special case $\mu =1$, equations (\ref{eq69}) and (\ref{eq71}) include
the well-know generating function equations for the standard Stirling
numbers of the second kind $S(m,l)$ (for example, see Eqs.(2.17) and (2.18)
in Ref.\cite{Charalambides}).

\end{document}